\documentclass[prl,aps,twocolumn,floats,showpacs]{revtex4}
\usepackage{graphicx}

\newcommand{\nin}{\noindent}

\newcommand{\be}{\begin{equation}}
\newcommand{\ee}{\end{equation}}
\newcommand{\bea}{\begin{eqnarray}}
\newcommand{\eea}{\end{eqnarray}}

\newcommand{\lb}{\left[}
\newcommand{\rb}{\right]}
\newcommand{\lp}{\left(}
\newcommand{\rp}{\right)}
\newcommand{\lf}{\left\{}
\newcommand{\rf}{\right\}}

\renewcommand{\H}{{\cal H}}
\renewcommand{\L}{{\cal L}}

\newcommand{\F}{{\cal F}}

\newcommand{\lsaw}{\lambda_{\rm ext}}
\newcommand{\ksaw}{k_{\rm ext}}

\newcommand{\ts}[1]{\textstyle{#1}}
\newcommand{\ds}[1]{\displaystyle{#1}}

\newcommand{\lch}{{l_{\rm ch}}}
\newcommand{\lfl}{{l_{\rm fl}}}

\newcommand{\mtot}{m_{\rm tot}}

\begin{document}
\title{Devil's staircase of incompressible electron states in a nanotube}
\author{Dmitry S. Novikov}
\email{dima@alum.mit.edu}
\affiliation{Department of Electrical Engineering and Department of Physics,
Princeton University, Princeton, NJ 08544}
\date{December 16, 2004}


\begin{abstract}

\nin
It is shown that a periodic potential applied to a nanotube can lock electrons into 
incompressible states.
Depending on whether electrons are weakly or tightly bound to the potential,
excitation gaps open up either due to  
the Bragg diffraction enhanced by the Tomonaga -- Luttinger correlations, 
or via pinning of the Wigner crystal.
Incompressible states can be detected in a Thouless pump setup, in which
a slowly moving periodic potential induces  quantized current,
with a possibility to pump on average a fraction of an electron per cycle
as a result of interactions.

\end{abstract}

\maketitle

\nin
A carbon nanotube (NT) is a strongly interacting electron system
known to be a host of many-body effects \cite{Nygard}, 
including Luttinger liquid behavior \cite{Stone}.
At half-filling the NT is effectively dilute, and a crossover from the
liquid to the 1d Wigner crystal \cite{WC} is expected \cite{Levitov'01}.

Here we suggest that an external periodic potential can be a probe 
of both crystallization and Luttinger correlations. 
We show that incompressible electron states arise when the 
electron number density $\bar\rho$ (relative to half--filling) is
{\it commensurate} with the potential period $\lsaw$:
\be \label{rho-alpha}
\bar \rho 
= {\mtot / \lsaw} \,,  \quad \mtot = 4m \,.
\ee
In Eq.~(\ref{rho-alpha}), $m$ is the number of the NT electrons of each 
of the four polarizations \cite{Dresselhaus} 
per period. 
To calculate excitation gaps we 
generalize the Pokrovsky--Talapov theory \cite{PT}
for the case of the four 
coupled fermion modes.

In the absence of interactions, Bragg diffraction 
on the potential
opens minigaps 
at integer density (\ref{rho-alpha}), 
$|m|=1,2,...$ \cite{Talyanskii'01}. 
Interactions profoundly change the spectrum, yielding 
a devil's staircase of incompressible states at {\it rational} $m=p/q$.
In such a state, the NT electron system is locked by the 
potential into a $q\lsaw$--periodic structure.
If detected, {e.g.} in a Thouless pump setup \cite{Talyanskii'01},
corresponding minigaps would provide a direct probe of interactions, 
with a possibility to map the devil's staircase by
pumping at fractions of the base frequency. 



One-dimensional interacting electrons are conventionally described 
by the Tomonaga - Luttinger liquid \cite{Stone}.
This hydrodynamic approach 
is valid 
in a small momentum shell near the Fermi points,
with excitations extended over the whole system.
%
%
%
Adequate description of crystallization  and commensurability 
requires
including
the {\it curvature} of the electronic dispersion
that becomes important at low density.
The curvature 
can yield crystallization or commensuration by
coupling charge and spin modes 
and by introducing a {\it length scale} 
into an otherwise scale-invariant Gaussian theory. 
In this work we treat both electron interactions and the curvature
of the dispersion non-perturbatively
by making use of the 
relativistic Dirac spectrum of a half--filled nanotube.
Curvature is controlled by the Dirac gap and is bosonized exactly
by virtue of the massive Thirring - sine-Gordon duality. \cite{Coleman,Haldane}
This enables us to study incompressible states both in the limit
of a narrow--gap Luttinger liquid and in that of the locked Wigner crystal.
%




Our course of action is to introduce the bosonized description for the NT electrons,
develop the phase soliton method 
and find excitation gaps from the renormalized 
sine--Gordon action, 
draw the phase diagram in the semiclassical limit, and comment on 
the experimental means to detect incompressible states.

{\it The model.---} 
Nanotube electrons in the forward scattering approximation \cite{fsa}
are described by the four flavors of Dirac fermions
$\psi_{\alpha} = \lp {\psi^R_{\alpha} \ \psi^L_{\alpha}} \rp^T$, 
$\alpha=1-4$,
whose interaction is written in terms of the smooth envelope
$\rho(x)=\sum_{\alpha = 1}^4\psi^+_{\alpha}(x)\psi_{\alpha}(x) $
of the total charge density.
The second-quantized Hamiltonian
$\H=\H_0 + \H_{\rm bs} + \H_{\rm ext}$,
where $\H_0$ is the massless Dirac Hamiltonian
\be \label{H0}
\H_0 = -i\hbar v\int 
\sum_{\alpha = 1}^4\psi^+_{\alpha}\sigma_3\partial_x\psi_{\alpha} dx 
+ \frac{1}{2}\sum_k\rho_{-k} V(k)\rho_{k} 
\ee
with the Coulomb interaction 
$V(k) = \frac{2e^2}{\varepsilon+1}\ln\lb 1+(ka)^{-2}\rb$
for a tube of radius $a$ 
placed on a substrate with the dielectric constant $\varepsilon$.
The curvature of the electron dispersion controlled by
the gap $2\Delta_0$ at half-filling introduces backscattering
at each NT Dirac point:
\be \label{Hbs}
\H_{\rm bs}  =  
\Delta_0 \int 
{\textstyle \sum_{\alpha = 1}^4} \psi^+_{\alpha}\sigma_1\psi_{\alpha} dx \ .
\ee
We emphasize that the backscattering  $\Delta_0$ is {\it not} the usual
interaction-induced $V(2k_F)$ term (which is undetectably small in metallic NTs), 
but rather is present at the single-particle level
\cite{Dresselhaus,curvature}:
Depending on the tube chirality, the bare gap $\Delta_0$ 
can be in the range $\Delta_0\lesssim$10\,meV to $\sim$0.5\,eV;
it
can also be controlled by the parallel magnetic field \cite{B-paral}.
Adding the periodic potential $U(x)$ and the chemical potential $\mu$
[$\mu=0$ at half-filling] results in
\bea
\label{Hext}
\H_{\rm ext} = \int \! dx \, \rho(x) \lf U(x)-\mu \rf \ ,\\
\label{SAW}
U(x) = A \cos \ksaw x \ , \quad \ksaw={2\pi / \lsaw} \ .
\eea
Qualitatively, our findings will be valid for any realistic potential
which justifies the simplest choice (\ref{SAW});
typically, $\lsaw \sim 0.1-1\,\mu$m.
The Hamiltonian ${\cal H}$
is U(4) invariant with respect to rotations in the fermion flavor space. 

Bosonization of the nanotube electrons 
$\psi_{\alpha} = {1\over \sqrt{2\pi a}}\, e^{i\Theta_{\alpha}}$
is exact even in the presence of (\ref{Hbs}). 
It maps the problem of the four interacting Dirac fermion modes onto 
the sine-Gordon model of the four coupled bose fields $\Theta_{\alpha}$.
We rotate \cite{Levitov'01,Talyanskii'01} 
to the basis of the charge mode $\theta^0$ and three neutral modes $\theta^a$,
in which case the charge density 
$\rho(x) = \textstyle{\frac{2}{\pi}}\, \partial_x \theta^0$, and 
the Gaussian action (bosonized $\H_0$) is diagonal
[$\hbar = v =1$]:
\be
\L_0 =  {1\over 2\pi} \int\! dx  
\lp
(\partial_{t} \theta^0)^2 - K\, (\partial_{x}\theta^0)^2 
+
\ts{\sum_{a=1}^3 \lp 
\partial_{\mu}\theta^a
\rp^2 }  \rp \, .
\label{L0}
\ee
The slow momentum dependence of the charge stiffness 
$K_k = 1 + 4\nu V(k)$, $\nu^{-1} = {\pi \hbar v}$,
is irrelevant, and we take it as constant 
$K\!\equiv\! K_{k\sim 1/\lch}$,  
$\lch$ being the size of the charged-mode soliton
(described below).
Assuming $\lch \sim l_s$, with the screening length $l_s \sim 1\, \mu$m,  
and using $e^2/\hbar v\simeq 2.7$, 
one estimates $K\simeq 40$ for 
for the stand-alone tube; $K\simeq 10$ 
if the tube is placed on a substrate
with 
dielectric constant $\varepsilon\simeq 10$.   
The logarithmic behavior of 
$V(k)$ 
underlies the fact 
that the Coulomb interaction $V(x)\propto 1/|x|$
is essentially {\it local} in one dimension, 
$\frac1{2} \int dx dx' \rho(x)V(x-x')\rho(x') 
\simeq \frac{\hbar v}{2\pi}\int dx K (\partial_x\theta^0)^2$.

The nonlinear part of the sine-Gordon Lagrangian [coming from $\H_{\rm bs}$]
reads 
\cite{Talyanskii'01,Levitov'01,fracnt} 
\bea
\L_{\rm bs} = -  \int\! dx \, g_0 
\F \lp
\theta^0 + 2\tilde\mu\ksaw x - 2\tilde A \sin \ksaw x ,\ \theta^a\rp 
\ ,
\label{LF}
\\
\label{f} 
\F \lp\theta^0, \,
\theta^a\rp
= \cos\theta^0 \cdot \ts{\prod_{a=1}^3 \cos\theta^a +  
\sin\theta^0 \cdot \prod_{a=1}^3 \sin\theta^a}  \ , 
\eea
where 
$g_0= {4 \Delta_0/\pi D a^2}$, 
and $D\simeq \hbar v/a$ is the 1d bandwidth. 
In Eq.~(\ref{LF}) we included the 
coupling (\ref{Hext}) to external fields
by shifting the charge mode 
$\theta^0 \to \theta^0 - \textstyle\frac2{\hbar v}\int^x K^{-1} 
(U-\mu)\, dx'$, 
with $\tilde\mu = \mu/(K\epsilon_0)$, $\tilde A = A/(K\epsilon_0)$,
$\epsilon_0=\hbar \ksaw v$.

In what follows, it is useful to 
first describe elementary excitations of the stand-alone tube, $U\equiv 0$.
As shown by Levitov and Tsvelik \cite{Levitov'01}, in the bosonized language
adding one electron corresponds to a  
{\it composite soliton} of both the charge and the flavor modes. 
In such a composite object, the charge mode changes by $\pi/2$
(adding unit charge) over the length $\lch$, 
whereas the neutral sector adds a particular SU(4) flavor to the
electron by means of the solitons of $\theta^a$
``switching'' by $\pm \pi/2$ 
on a shorter scale 
$\lfl \sim K^{-1/2}\lch$,
right in the middle of the charge soliton.
The composite soliton is a unit charge configuration
of the minimal energy, obtained by optimizing the 
action $\L_0 + \L_{\rm bs}$ in the limit of large Coulomb repulsion $K\gg 1$.
Technically, such an optimization results in the soft neutral modes $\theta^a$
adjusting to create the effective potential 
$\bar \F(\theta^0) = \min_{\{ \theta^a\} } \F (\theta^a , \ \theta^0)
\sim \cos 4\theta^0$ for the stiff charge mode $\theta^0$.

In the absence of external potential, $U\equiv 0$,
the system $\L_0 + \L_{\rm bs}$ 
describes the Wigner crystal - Luttinger liquid crossover.
In particular, raising the chemical potential from
half-filling 
[band insulator, or the ``Dirac vacuum'']
to just above the charge gap 
produces the train of weakly overlapping
composite solitons described above [$\bar\rho\lfl \ll 1$, ``Wigner crystal''],
in which overlapping charge solitons 
maintain a quasi-long-range order. 
Further increase of 
$\mu$ leads to 
strongly overlapping solitons rendering the nonlinear term (\ref{LF}) irrelevant
[$\bar\rho\lfl\gg 1$, Luttinger liquid].
We emphasize that the crystal - liquid crossover occurs due to the
{\it finite} soliton size $\lfl\propto \Delta_0^{-\zeta}$ 
that scales inversely with the {curvature} of the dispersion.
Curvature is also responsible for binding flavor to charge through the term 
(\ref{LF}).
The $U\equiv 0$ system is compressible \cite{Levitov'01}.

Periodic potential locks electrons into incompressible states.
Technically, the term (\ref{LF}) becomes relevant whenever the density 
$2\tilde\mu = m$ [Eq.~(\ref{rho-alpha})] is integer \cite{Talyanskii'01}.
When $m$ is a simple fraction, $2\tilde\mu = p/q$, to identify incompressible states 
one utilizes the phase soliton method \cite{PT}.
We generalize it for the case of the four nanotube modes   
by expanding in powers of the coupling $g_0$ as 
$\theta^j = 
\bar \theta^j + g_0{\theta^j}^{(1)} + ... + g_0^n{\theta^j}^{(n)} + ... $,
$j=0, a$, 
and finding the effective Lagrangian 
$\L_{m}[\bar \theta^0, \bar \theta^a]$ 
for the {\it phase} modes $\bar\theta^0$ and $\bar\theta^a$, $a=1-3$, 
to the lowest order in $g_0$. 
The phase modes $\bar \theta^0, \bar \theta^a$ 
are constant in the commensurate phase, whereas an
excitation is a {\it composite phase soliton}, in which the
phase fields $\bar \theta^0(x), \bar \theta^a(x)$ describe a slow 
deformation of the regular commensurate configuration. 
Excitation gap is given by the energy of the 
composite soliton, renormalized by quantum fluctuations.

We note that it is the curvature $\propto\Delta_0$ that yields 
incompressible states.
Same is true for the non-interacting electrons: 
When $\Delta_0=0$, the external potential is gauged away from 
the Hamiltonian $-i\hbar v \sigma_3\partial_x + U(x)$.

The composite phase soliton is a result of optimization of the 
corresponding effective action $\L_{m}[\bar \theta^0, \bar \theta^a]$
(examples of which are given below). 
This problem will be similar 
to the $U\equiv 0$ case described above: 
When interaction is strong, $K\gg 1$, the neutral phase modes 
$\bar\theta^a$ adjust [on the scale $\bar\lfl$]
to create an effective potential for the charged 
phase mode $\bar\theta^0$.
The crucial difference from the former case is that 
this optimization 
will qualitatively depend
on whether the system is in the regime of 
the Luttinger liquid or in that of the Wigner crystal. In this sense,
periodic potential naturally distinguishes between the opposite sides of the 
crystal - liquid crossover, by bringing about the additional
{\it length scale}, its period $\lsaw$.
Technically, the saddle point of $\L_m$ will depend on   
whether the neutral modes 
are adiabatic or fast on the length scale on 
which the effective potential (\ref{LF}) changes appreciably.
Below we consider both cases separately.

{\it Bragg diffraction in a four-flavor Luttinger liquid.---}
In the adiabatic limit $\bar\lfl \gg \lsaw$ of extended flavor excitations,
{\it exchange} is important:
The system
(correlated over many $\lsaw$) ``knows'' that it is comprised of particles
of the four different flavors. This limit is 
connected to the non-interacting case, where particles repel 
only due to the Pauli principle. 
For integer density $m$, averaging (\ref{LF}) over the period $\lsaw$ 
reduces the problem to the $U\equiv 0$ case 
with $\L_m = \L_0 + \L_{\rm bs}$, 
$g_0 \to g_0 J_m(2\tilde A)$, 
yielding the $K\gg 1$ minigaps 
$\Delta_m \simeq K^{1/2} D^{1/5} | \Delta_m^{(0)}|^{4/5}$
that are enhanced by the quantum fluctuations compared to the noninteracting
values $\Delta_m^{(0)} = 2\Delta_0 J_m(2\tilde A)$ \cite{Talyanskii'01}.

The fractional $m=\frac12$ case is considered in detail in Ref.~\onlinecite{fracnt}.
A somewhat lengthy calculation yields the effective action of the form
$\L_{1/2} = \L_0 
+ 
v(\tilde A) \sum_{a} \cos 2\bar\theta^a \cos 2\bar\theta^0 
- u(\tilde A) \sum_{a>b} \cos 2\bar\theta^a \cos 2\bar\theta^b
$. 
We find the {\it charge} excitation gap 
$\Delta_{1/2} \propto 
{K-1\over \sqrt{K}} \, D 
\lp {\Delta_0 \over \epsilon_0}\rp^2$
(that vanishes in the noninteracting limit $K\to 1$ 
in accord with the Bloch theory),
whereas {\it flavor} 
excitations are governed by the SU(4)$\simeq$O(6) Gross - Neveu
Lagrangian \cite{GN} that can be written in terms of six Majorana fermions
$\chi \sim e^{i\theta}$,
$\L_{\rm GN} 
=  i\bar\chi_j \gamma_{\mu}\partial_{\mu}\chi_j - 
u(\tilde A) (\bar \chi_j \chi_j)(\bar \chi_{j'} \chi_{j'})$.
Although the charge gap $\Delta_{1/2}$ vanishes for certain values of the potential
amplitude (zeroes of $v(\tilde A)$ given in Ref.~\onlinecite{fracnt}), 
the excitation gap never closes due to flavor,
since the Gross - Neveu coupling $u(\tilde A) \neq 0$.

{\it Locked Wigner crystal.---}
Below we focus on the 
opposite limit $\bar\lfl \ll \lsaw$, in which the charge
mode dominates, whereas exchange (overlap of the flavor solitons) is
relatively unimportant. 
We will show that the effective description in this regime is that 
of a {\it single} mode of fermions with the density (\ref{rho-alpha}),
locked by the external potential.
Physically this happens since when the electron wavefunctions (represented 
by solitons) are {localized} on the scale $\lsaw$, 
Coulomb repulsion wins over exchange.
Naturally, in this case 
the appropriate saddle point of $\L_0 + \L_{\rm bs}$ 
is equivalent to
the semi-classical 
description, with the WKB condition $\bar\lfl \ll \lsaw$. 
Technically, 
the neutral mode ``switching'' produces the effective charge mode potential
with a quarter-period, 
similarly to the $U\equiv 0$ case above:
$g_0 \F ( \theta^0 + m\ksaw x - 2\tilde A \sin \ksaw x ,\ \theta^a)
\sim g \cos ( 4\theta^0 + \mtot\ksaw x -8\tilde A \sin \ksaw x)$. 
This potential  now depends
on the {\it total density} $\mtot$, Eq.~(\ref{rho-alpha}). 

To obtain the renormalized coupling $g$ we utilize adiabaticity of $U(x)$,
$\bar\lfl \ll \lsaw$.  
The renormalization group 
produces the effective coupling $g$ 
on the scales $a < l < \lfl$ that appear
``microscopic'' for the external potential.
Thus the neutral modes can be integrated out, yielding 
both $\lfl$ and $g$ 
independent of the shape of the potential $U(x)$.
The problem 
becomes similar to the $U\equiv 0$ case.
The coupling $g$ 
follows from the 
scaling $g \simeq g_0 (\lfl/a)^{-3/4}$ and 
self-consistency $g(\lfl)\sim 1/\lfl^2$. 
\cite{Talyanskii'01,Levitov'01,fracnt}

{\it Re-fermionization.---}
Let us now complete our single-mode description by 
mapping the system in the limit $\bar\lfl \ll \lsaw$ onto the problem 
of spinless Dirac fermions in the external potential (\ref{SAW}).
We introduce the displacement field 
$\Theta=2\theta^0$  
for the total density
$\rho\equiv \frac1{\pi} \partial_x \Theta$.
To preserve commutation relations,
we rescale the canonical momentum 
$\Pi_{\Theta}=\frac12 \Pi_{\theta^0}$. 
Changing variables in the Lagrangian 
$\frac{1}{2\pi}\lf K(\partial_x\theta^0)^2 + g\cos(4\theta^0 + ...)\rf$, 
we obtain the effective Lagrangian for the charge mode $\Theta$,
\be
\matrix{
\L_{\rm eff}[\Theta] = \ds{\hbar v'\over\pi} 
\int\! dx  
\lf
\frac1{2v'^2} \lp \partial_t  \Theta\rp^2 
- \frac{K'}2\, \lp\partial_{x}\Theta\rp^2 
\right. \cr \left.
- g \cos(2\Theta+\mtot\ksaw x - 2\tilde A'\sin\ksaw x)
\vphantom{\ds{\int {1\over v^2}}}\rf 
}
\label{L-Theta}
\ee
with the 
rescaled parameters 
$v' \equiv 4v$, $K' \equiv {K/16}$, and 
$\tilde A' \equiv {A/ K'\hbar \ksaw v'} = 4\tilde A$.
The velocity quadrupling corresponds to counting incoming fermions
regardless of their (four) flavors. 
With the same accuracy that allowed us to neglect 
the flavor sector, 
the rescaled value $K'\approx 1 + {(K-1)/16} \equiv 1 + \nu' V(q)$
is by definition the charge stiffness for the spinless Dirac fermions
of velocity $v'$, with the density of states $\nu'={1/\pi\hbar v'}$.
The external fields $U$ and $\mu$ are not rescaled.
%
%
Introducing the Dirac spinors 
$\Psi={1\over \sqrt{2\pi a'}} e^{i\Theta}$ relative to the new 
cutoff $a'\sim \lfl$, 
we obtain the 
effective Hamiltonian 
\be 
\matrix{
\H_{\rm eff}[\Psi] = \ds{\int}\! dx \, 
\Psi^+ \lf -i\hbar v'\sigma_3 \partial_x  + \Delta' \sigma_1 
+ U(x) \right. \cr \left.
- \mu \rf \Psi 
+ \ds{\frac12}\sum_k\rho_{-k}V(k)\rho_{k}
}
\label{H-strong}
\ee
for the fictitious 
spinless Dirac fermions of unit charge, 
with the number density 
$\rho(x) = \Psi^+\Psi$,
coupled to the external fields in the natural way, Eq.~(\ref{Hext}). 
The fermions $\Psi$ represent the original NT electrons 
traced over the flavor states.
The effective gap $\Delta'$ in Eq.~(\ref{H-strong}) 
corresponds
to the renormalized coupling $g$
entering the Lagrangian 
(\ref{L-Theta})
by $g\simeq {\Delta' \over \hbar v' a'}$, 
yielding
$\Delta' \simeq {\hbar v' \over \lfl} \simeq D^{1/5} \Delta_0^{4/5}$.
 
The (charge) excitation gaps are now estimated via the 
phase soliton approach \cite{PT}
for the single-mode system (\ref{L-Theta}) and (\ref{H-strong}),
yielding the devil's staircase.
For integer $\mtot=\pm1, \pm2, ...$, 
in the classical limit $K\to \infty$, we find
\be
\label{minigaps-ren-4}
\Delta_{m} \simeq K^{1/2} 
|J_{\mtot}(2\tilde A')|^{1/2}
D^{1/5}\Delta_0^{4/5} \ .
\ee
The effective density quadrupling $m\to \mtot$ in (\ref{minigaps-ren-4})
is due to strong interactions: Electrons of all flavors avoid each other,
their ground state being a Slater determinant of the quadrupled size.

\begin{figure}[t]
\centerline{
\includegraphics[width=1.8in]{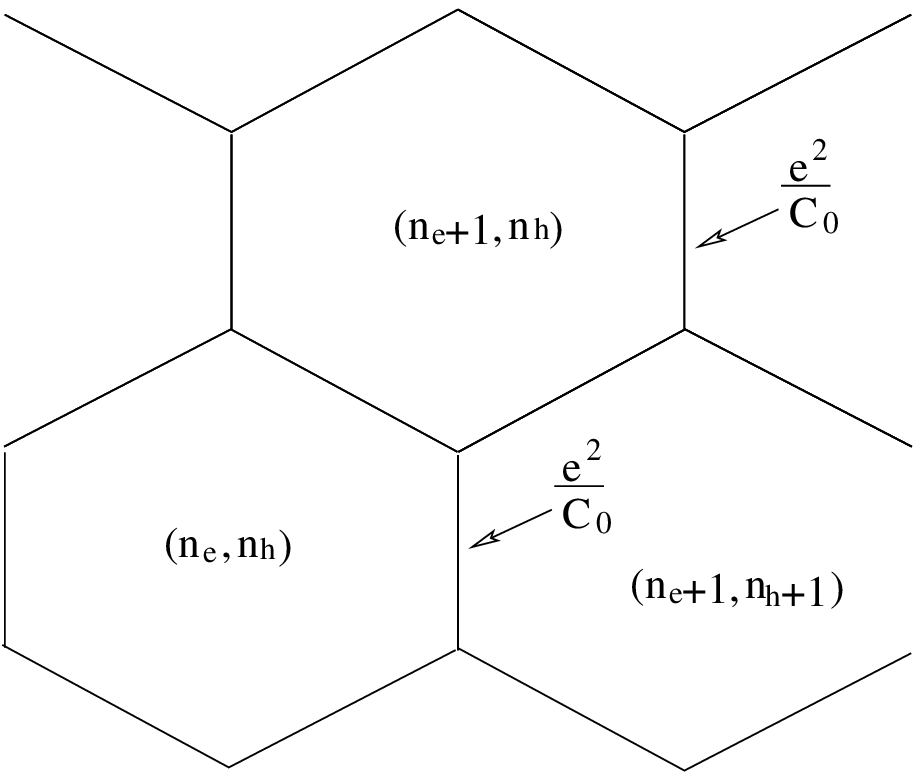}
\includegraphics[width=1.7in,height=1.5in]{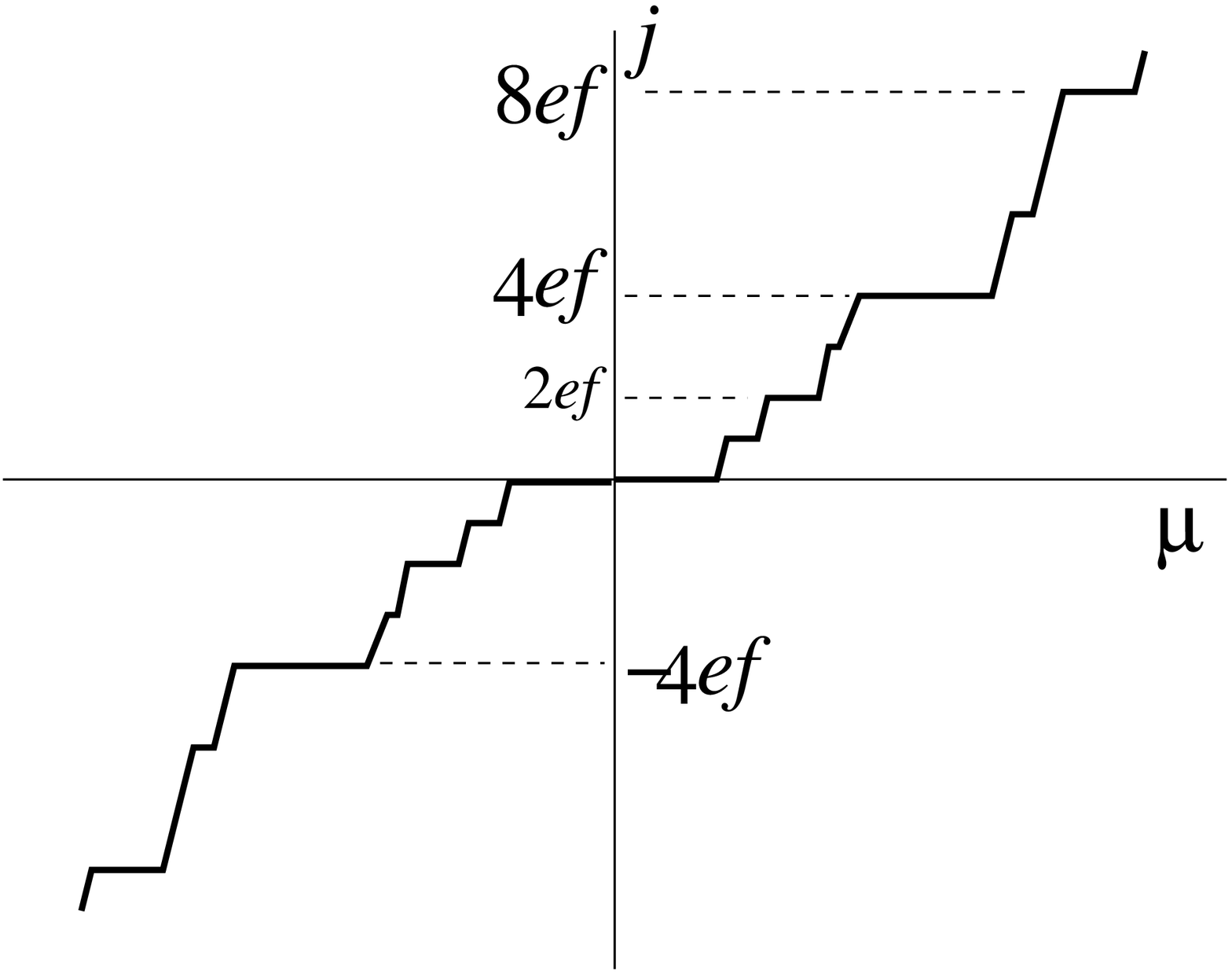}
}
\caption[]{
{\it Left:} 
Phase diagram in the $(A, \mu)$ plane 
according to the model (\ref{hamiltonian-cl}).
In the limit  $n_e, n_h \gg 1$ 
the period in $A$ becomes $e^2/C_1$.
{\it Right:} 
Mapping the devil's staircase in the Thouless pump setup.
In addition to the integer-$m$ plateaus, the ones with fractional $m$ appear
due to electron interactions. 
}
\label{fig:phdiag}
\end{figure}

{\it Phase diagram.---}
%
The Hamiltonian (\ref{H-strong}) in the limit $K\gg 1$,
coarse-grained beyond $\lfl$, describes the classical 
chain of $n_e$ electrons and $n_h$ holes per period, $\mtot=n_e-n_h$.
Their positions $x_i$ in the minima and $y_j$ in the maxima of $U(x)$,
as well as the characteristic Coulomb blockade hexagons
labeled by ($n_e$, $n_h$) in Fig.~\ref{fig:phdiag},
are obtained by minimizing $E_{\rm cl}(A)/N - \mu(n_e-n_h)$, where
\begin{eqnarray}
\matrix{\ds
E_{\rm cl} = 
\sum
\lb \Delta' + U(x_i) \rb 
\ + \ 
\sum
\lb \Delta' - U(y_j) \rb 
\cr \ds
+ \sum_{i>i'} V(x_i-x_{i'})
+ \sum_{j>j'} V(y_j-y_{j'})
- \sum_{i,j} V(x_i-y_j) \ \ 
}
\label{hamiltonian-cl}
\end{eqnarray}
for a finite tube of the length $L=N\lsaw$.
In Fig.~\ref{fig:phdiag}, 
the hexagon size is controlled by the NT capacitance per period, 
$C_0\simeq {\lsaw\over 2\ln(l_s/\lsaw)}$. 
For $n_e, n_h \gg 1$, the Thomas - Fermi approximation
$\rho(x)\approx \rho_{\rm TF}(x) \propto U(x)$
estimates the interaction energy 
inside each half-period (``quantum dot'') as
${e^2\over 2} \int
\! dx dx'\, 
{\rho_{\rm TF}(x)\rho_{\rm TF}(x') \over |x-x'|} 
\equiv {n_e^2 e^2\over 2C_1}$,
where the ``dot capacitance'' 
$C_1 \simeq {\lsaw \over 2\pi \ln (\lsaw/a')}$. 
In this limit, all the hexagon domains of 
the phase diagram 
become identical with the width
${e^2/C_1}$. 
Remarkably, this period asymptotically follows from Eq.~(\ref{minigaps-ren-4})
with $\tilde A' 
\approx
{\pi\over 2} \cdot {A\over e^2/C_1}$ [$K\equiv K_{1/\lsaw} \approx K-1$],
using the Bessel function zeroes 
$2\tilde {A'}_{\mtot}^{(n)} \approx {3\pi\over 4} + {\pi \mtot \over 2} + \pi n$, 
$n={\rm min}\{ n_e, n_h \}$.
The borders between the domains correspond to the incompressible states
with {fractional} $n_e$ and $n_h$ (not shown). 

{\it Quantized current.---}
The devil's staircase can be mapped in 
the Thouless pump setup \cite{Talyanskii'01}.
With the Fermi level in the gap,
a slowly moving wave 
$U(x-st)$ with the frequency  $f=s/\lsaw$ 
will generate the quantized current 
$j = \mtot ef$. 
The moving potential can be created by
gating, optical methods, or acoustic field.
Estimated gap values $\Delta_m$ in the meV range ensure adiabaticity 
leading to current quantization and possible metrological applications. 
Novel incompressible states with {fractional} $m$ would correspond  
to the current quantized in 
the  {\it fractions} of $4ef$, 
as illustrated in Fig.~\ref{fig:phdiag}.  
Current changes sign at half-filling
due to the Dirac symmetry. 

{\it Conclusions.---}
We demonstrated that coupling to a periodic potential
results in the devil's staircase of incompressible electron states.
Excitation gaps are found
in the limit of the narrow-gap Luttinger liquid and in that of the  
Wigner crystal, by selecting the 
saddle point of the nonlinear action of the four bosonic modes.
When the Coulomb interaction dominates, 
the system behaves as a single 
fermion mode with renormalized mass and velocity. 
Control over the NT gap, the screening length, and the parameters of the  
potential makes this setup
a unique probe of the Luttinger liquid - Wigner crystal crossover and 
of commensuration effects in 1d.
Novel effect of adiabatic pumping at fractions of base frequency is linked to
the interaction-induced incompressible states.
The effective single-mode description could also rationalize
recent manifestations of Wigner crystallization in transport,
such as the $e^2/h$ steps in conductance \cite{Matveev0.7}.


This work was supported by NSF MRSEC grant DMR 02-13706.


\end{document}